\documentclass[12pt, a4paper] {article}

\usepackage{amsfonts}
\usepackage{verbatim}
\usepackage{color}

\begin{document}

\begin{center}

{\bf {\large New observations in the BRST analysis of dynamical 
             non-Abelian 2-form gauge theory}}

\vskip 2.5 cm

{\bf{ R. Kumar$^{(a)}$, R. P. Malik$^{(a,b)}$}}\\
{\it $^{(a)}$Department of Physics, Centre of Advanced Studies,}\\
{\it Banaras Hindu University, Varanasi - 221 005, (U. P.), India}\\

\vskip 0.1cm

{\bf and}\\

\vskip 0.1cm

{\it $^{(b)}$DST Centre for Interdisciplinary Mathematical Sciences,}\\
{\it Faculty of Science, Banaras Hindu University, Varanasi - 221 005, India}\\

\end{center}

\vskip 2.5 cm

\noindent

\noindent
{\bf Abstract:} We generalize the {\it usual} gauge transformations connected with the 1-form gauge potential to the Becchi-Rouet-Stora-Tyutin (BRST) and anti-BRST symmetry transformations for the four $(3+1)$-dimensional (4D) topologically massive non-Abelian gauge theory
that incorporates the famous $(B\wedge F)$ term where there is an explicit topological coupling between 1-form and 2-form gauge fields. A novel feature of our present investigation is the observation that the (anti-)BRST symmetry transformations for the auxiliary 1-form field $(K_\mu)$ and 2-form gauge potential $(B_{0i})$ are {\it not} generated by the (anti-)BRST
charges that are derived by exploiting {\it all} the relevant (anti-)BRST symmetry transformations corresponding to {\it all} the fields of the present theory. This 
observation is a new result because it is drastically different from the application of the
BRST formalism to (non-)Abelian 1-form and Abelian 2-form as well as 3-form gauge theories.\\ 



\newpage

\section{Introduction}

In recent years, there has been a great deal of interest in the study of 
higher $p$-form ($p = 2, 3, 4...$) gauge theories because of their relevance
in the context of (super)string theories and related extended objects
(see, e.g. [1,2]). The merging of the 1-form and 2-form gauge fields has
provided us with the topological massive gauge theories in 4D. In such
(non-)Abelian theories, the 1-form gauge field acquires mass in a very 
natural fashion [3]. As a consequence, it provides an alternative to the
method of mass generation by Higgs mechanism in the context of
standard model of high energy physics.

In view of the fact that the Higgs particles of the standard model have not 
yet been observed experimentally, the above 4D topologically
massive (non-)Abelian theories [3-7] have attracted a renewed interest in the
recent past. In this context, it is pertinent to point out that we have studied
the 4D topologically massive {\it Abelian}  2-form gauge theories within the frameworks 
of superfield and BRST formalisms [8,9] and derived the absolutely anticommuting (anti-)BRST
symmetry transformations. We have also considered the dynamical non-Abelian 2-form theory
within the superfield scheme [10] where we have exploited its ``scalar'' and ``vector''
gauge symmetry transformations to derive the proper (anti-)BRST symmetry transformations.
In a very recent publication [11], we have derived the coupled Lagrangian densities
that respect the above off-shell nilpotent and absolutely anticommuting
(anti-)BRST transformations corresponding to the ``scalar''
gauge symmetry.

The purpose of our present Letter is to derive the off-shell nilpotent symmetry generators
(i.e. conserved charges) for the above off-shell nilpotent (anti-)BRST symmetry transformations
[11] and derive their corresponding BRST algebra. One of the novel observations of our present endeavor is the finding that the generators of the above nilpotent symmetry transformations
do {\it not} generate the symmetry transformations corresponding to the auxiliary vector
field $K_\mu$ and the $B_{0i}$ component of the anti-symmetric tensor gauge field $B_{\mu\nu}$.
We provide the possible 
reasons behind this novel observation in the language of the constraints of the theory. We would like to lay emphasis on the fact that our present novel observation, in the
context of the dynamical non-Abelian 2-form theory, is a new result and it is
drastically different from the
application of the BRST formalism to (non-)Abelian 1-form [12,13] and Abelian
2-form as well as 3-form gauge theories in 4D [8,9,14].

Our present paper is organized as follows. In second section, we discuss the local gauge symmetry transformations and their generator
corresponding to the 1-form non-Abelian gauge field. Our section three is devoted to the up-gradation of the above gauge symmetry transformations
to the off-shell nilpotent BRST symmetries and derivation of 
the corresponding conserved charge. Section four deals with the anti-BRST symmetries 
and their generator. The ghost symmetries and corresponding generator are discussed and derived in our section five. We also deduce BRST
algebra, in this section, in a simple manner. 
Finally, in section six, we make some concluding remarks.

\section{Preliminaries: usual local gauge symmetry transformations and their generator}

We begin with the Lagrangian density of the 
4D topologically massive non-Abelian gauge theory\footnote{We adopt the convention and notations such that the background 4D
spacetime metric is flat with signature $(+1, -1, -1, -1)$ so that 
the dot product between two non-null vectors: 
$A\cdot B = A_\mu B^\mu = A_0 B_0 - A_i B_i$. Here the Greek indices 
$\mu, \nu, \eta,... = 0, 1, 2, 3$ and the Latin indices $i, j, k,... = 1, 2, 3$. In the algebraic space, for the sake of brevity, we choose dot and cross products between two vectors
as: $P \cdot Q = P^a Q^a$ 
and $(P \times Q)^a = f^{abc} P^b Q^c$ where $a, b, c,... = 1,2,,.... (N^2-1)$ for 
the $SU(N)$ Lie algebra.} that incorporates the topological mass parameter $m$ through
the celebrated $(B \wedge F)$ term. This is given by  [6,7]
\begin{eqnarray}
{\cal L}_0 = - \; \frac {1}{4} \;F^{\mu\nu} \cdot F_{\mu\nu} + \; \frac {1}{12}\; H^{\mu\nu\eta} \cdot H_{\mu\nu\eta} 
+ \frac {m}{4}\;\varepsilon^{\mu\nu\eta\kappa}\; B_{\mu\nu} \cdot F_{\eta\kappa}, 
\end{eqnarray}
where the 2-form curvature $F^{(2)} = d A^{(1)} + i\; A^{(1)} \wedge A^{(1)}$ defines the curvature tensor 
$F_{\mu\nu} =  \partial_\mu A_\nu - \partial_\nu A_\mu - (A_\mu \times A_\nu)$. In the above, the
2-form
$F^{(2)} = \frac {1}{2!}\;(dx^\mu \wedge dx^\nu) \; F_{\mu\nu}$ and the 1-form $A^{(1)} = dx^\mu A_\mu$ define the $SU(N)$ 
valued curvature tensor $F_{\mu\nu}$ and 
gauge potential $A_\mu$, respectively. Similarly, the 3-form 
$H^{(3)} = \frac {1}{3!}\;(dx^\mu \wedge dx^\nu \wedge dx^\eta)\; H_{\mu\nu\eta}$
defines the curvature tensor 
\begin{eqnarray}
&& H_{\mu\nu\eta} = (\partial_\mu B_{\nu\eta} + \partial_\nu B_{\eta\mu} + \partial_\eta 
B_{\mu\nu})
- \bigl [(A_\mu \times B_{\nu\eta}) + (A_\nu \times B_{\eta\mu}) \nonumber\\
&&+ (A_\eta \times B_{\mu\nu}) \bigr ] - \bigl [(K_\mu \times F_{\nu\eta}) 
+ (K_\nu \times F_{\eta\mu}) + (K_\eta \times F_{\mu\nu}) \bigr ],
\end{eqnarray}
in terms of the compensating non-Abelian
1-form $(K^{(1)} = dx^\mu K_\mu \cdot T)$ auxiliary field $K_\mu = K_\mu \cdot T$, the 
non-Abelian 2-form [$B^{(2)} = \frac {1}{2!}\; (dx^\mu \wedge dx^\nu)\; B_{\mu\nu} \cdot T$] 
gauge potential $B_{\mu\nu} = B_{\mu\nu} \cdot T$ 
and the non-Abelian 2-form curvature tensor $F_{\mu\nu} = F_{\mu\nu} \cdot T$ 
for the non-Abelian 1-form gauge field $A_\mu = A_\mu \cdot T$. Here
the $SU(N)$ generators $T^a$ satisfy the Lie-algebra 
$[T^a, \; T^b] = i\; f^{abc} \; T^c$ where $f^{abc}$ are the structure constants that 
have been chosen to be totally antisymmetric in indices $a, b,$ and $c$ 
for the semi-simple Lie group $SU(N)$ [13]. The above Lagrangian density (1) respects 
($\delta_{gt} {\cal L}_0 = 0$) the usual 
local gauge symmetry transformations ($\delta_{gt}$) 
corresponding to the 1-form gauge field as [6,7]
\begin{eqnarray}
&&\delta_{gt} A_\mu = D_\mu \Omega , 
\quad \delta_{gt} B_{\mu\nu} = - (B_{\mu\nu} \times \Omega ), 
\quad \delta_{gt} K_\mu = - (K_\mu \times \Omega  ), \nonumber\\
&&\delta_{gt} F_{\mu\nu} = - (F_{\mu\nu} \times \Omega  ), 
\;\;\qquad \;\;\;\delta_{gt}H_{\mu\nu\eta}= - (H_{\mu\nu\eta} \times \Omega ),
\end{eqnarray}
where $\Omega = \Omega \cdot T \equiv \Omega^a \;T^a$ is the $SU(N)$ valued infinitesimal ``scalar" gauge parameter\footnote{In addition to the local ``scalar''
gauge symmetry transformations (3), there also exists the ``vector'' gauge symmetry
transformations ($\delta_v$): $\delta_v A_\mu = 0, \delta_v F_{\mu\nu} = 0, \delta_v K_\mu = - \Lambda_\mu,   
\delta_v B_{\mu\nu} = - (D_\mu \Lambda_\nu - D_\nu \Lambda_\mu),  
\delta_v H_{\mu\nu\eta} = 0$ such that 
$\delta_v {\cal L}_0  = - (m/2)\; \partial_\mu [\varepsilon^{\mu\nu\eta\kappa}  
\Lambda_\nu \cdot F_{\eta\kappa}]$ where
$\Lambda_\mu = \Lambda_\mu \cdot T$ is an infinitesimal vector gauge parameter [6,7]. 
As a consequence, the action of the theory remains invariant under the ``vector'' gauge transformations 
$\delta_v$.} and the covariant derivative
$D_\mu \Omega = \partial_\mu \Omega - (A_\mu \times \Omega)$.

According to the Noether's theorem, the above infinitesimal continuous symmetry 
transformations lead to the following conserved current
\begin{eqnarray}
J^\mu_{gt} &=& \Big[\frac {m}{2} \; \varepsilon^{\mu\nu\eta\kappa} B_{\eta\kappa} - F^{\mu\nu}
- (H^{\mu\nu\eta} \times K_\eta) \Big] \cdot (D_\nu \Omega )  \nonumber\\
&-& \frac{1}{2} \; (H^{\mu\nu\eta} \times B_{\nu\eta})\cdot \Omega. 
\end{eqnarray}
To prove the conservation law of the above current, it is convenient to re-express
the above Noether current as given below
\begin{eqnarray}
J^\mu_{gt} &=& \;\partial_\nu \Big [ \frac {m}{2}  \varepsilon^{\mu\nu\eta\kappa}  B_{\eta\kappa}\cdot \Omega 
- (H^{\mu\nu\eta} \times K_\eta) \cdot \Omega - F^{\mu\nu} \cdot \Omega \Big ] \nonumber\\
& +& D_\nu \Big [ F^{\mu\nu} + (H^{\mu\nu\eta} \times K_\eta) 
- \frac {m}{2} \varepsilon^{\mu\nu\eta\kappa} B_{\eta\kappa} \Big] \cdot \Omega \nonumber\\
&-& \frac{1}{2} \; (H^{\mu\nu\eta} \times B_{\nu\eta})\cdot \Omega.
\end{eqnarray}
It can be checked that $\partial_\mu J^\mu_{gt} = 0$ if we use the following Euler-Lagrange equations of motion 
derived from the starting Lagrangian density (1):
\begin{eqnarray}
&&D_\mu \Big[F^{\mu\nu}  + (H^{\mu\nu\eta} \times K_\eta) - \frac {m}{2}\; \varepsilon^{\mu\nu\eta\kappa} 
\; B_{\eta\kappa} \Big]
= -\; \frac {1}{2} \;(H^{\nu\eta\kappa} \times B_{\eta\kappa}), \nonumber\\
&&D_\mu H^{\mu\nu\eta} = \frac {m}{2} \; \varepsilon^{\nu\eta\rho\sigma} \; F_{\rho\sigma}, \qquad 
\quad(H^{\mu\nu\eta} \times F_{\nu\eta}) = 0.
\end{eqnarray}
The above conserved current leads to the derivation of the conserved charge that turns out to be the generator 
of {\it a part} of  the gauge transformations (3). To corroborate the above statement, it can be checked that
(5) leads to the derivation of the generator 
($Q_{(gt)} = \int d^3 x J^0_{gt}$) of gauge transformations, as 
\begin{eqnarray}
Q_{(gt)} &=& \int d^3x \Big [\Big \{\frac {m}{2} \; \varepsilon^{0ijk} \;B_{jk} - F^{0i} 
- (H^{0ij} \times K_j)\Big \}\cdot (D_i \Omega ) \nonumber\\
&-&  \frac{1}{2} \; (H^{0ij} \times B_{ij}) \cdot \Omega  \Big ].
\end{eqnarray}
The above generator, however, generates only the following 
local and infinitesimal local gauge symmetry transformations of (3), namely;
\begin{eqnarray}
&& \delta_{gt} A_i(x) = - \; i \; [A_i(x), \; Q_{(gt)} ] =  \;D_i \;\Omega \;(x) , \nonumber\\
&& \delta_{gt} B_{ij}(x) = - \; i \; [B_{ij}(x), \; Q_{(gt)} ] = 
\;- \;(B_{ij} \times \Omega )\;(x).
\end{eqnarray}
We conclude that $Q_{(gt)}$ is {\it not} a full generator for
the transformations (3).

We wrap up this section with a couple of remarks. First, the auxiliary field $K_\mu$ leads to 
the constraint equation of motion $(H^{\mu\nu\eta} \times F_{\nu\eta}) = 0$ which can {\it not} be easily satisfied. 
However, we know that the Maurer-Cartan equation $F^{(2)} = d A^{(1)} + i\; (A^{(1)} \wedge A^{(1)})$ [which 
defines the curvature tensor $F_{\mu\nu} =  \partial_\mu A_\nu - \partial_\nu A_\mu - (A_\mu \times A_\nu)$ for
the 1-form gauge field] has a solution $A^{(1)}  = - i U d U^{-1} $ (where $U$ is an $SU(N)$ valued transformation 
function) such that the zero curvature condition $F_{\mu\nu} = 0$ is very naturally obtained\footnote{This feature 
is one of the key requirements of an integrable system (see, e.g. [15]).}. Second, the gauge transformations (3)
can be generalized to BRST and anti-BRST symmetry transformations that lead to the derivation of
generators that are more general than $Q_{(gt)}$. This is what precisely we do in our next sections.

\section{BRST symmetries and their generator}

The starting Lagrangian density (1) can be generalized to the BRST invariant Lagrangian density 
that incorporates the gauge-fixing and Faddeev-Popov ghost terms (in the Feynman gauge) as given below [11]
\begin{eqnarray}
{\cal L}_b &=& - \; \frac {1}{4} \;F^{\mu\nu} \cdot F_{\mu\nu} + \; \frac {1}{12}\; H^{\mu\nu\eta} \cdot H_{\mu\nu\eta} 
+ \frac {m}{4}\;\varepsilon^{\mu\nu\eta\kappa}\; B_{\mu\nu} \cdot F_{\eta\kappa} \nonumber\\
&+& B \cdot (\partial_\mu A^\mu) + \frac {1}{2}\; (B \cdot B + \bar B \cdot \bar B ) - i \; \partial_\mu \bar C \cdot D^\mu C.
\end{eqnarray}
The above Lagrangian density respects (i.e. $s_b {\cal L}_b = \partial_\mu [B \cdot D^\mu C]$) the following off-shell 
nilpotent ($s^2_b = 0$) BRST transformations ($s_b$) [11]
\begin{eqnarray}
&& s_b A_\mu = D_\mu C, \quad s_b C = \frac{1}{2} \;(C \times C), \quad s_b \bar C = i B, 
\quad s_b B = 0,  \nonumber\\ 
&& s_b  \bar B = - (\bar B \times C), \quad s_b  F_{\mu\nu} = - (F_{\mu\nu} \times C), 
\quad s_b K_\mu = - (K_\mu \times C), \nonumber\\
&& s_b H_{\mu\nu\eta} = - (H_{\mu\nu\eta} \times C), \;\;\qquad \;\;s_b B_{\mu\nu} = - (B_{\mu\nu} \times C),
\end{eqnarray}
where fermionic ($C^2 = \bar C^2 = 0, C\bar C + \bar C C = 0, etc. $) (anti-)ghost fields $(\bar C)C$ 
are required for the unitarity and $(B, \bar B)$ are the Nakanishi-Lautrup type auxiliary fields that 
satisfy the Curci-Ferrari (CF) restriction $[B + \bar B = - i (C \times \bar C)]$ which can be derived 
by exploiting the equations of motion (see, section 4 below). The above transformations lead to the following 
Noether current 
\begin{eqnarray}
J^\mu_b &=& - F^{\mu\nu} \cdot D_\nu C + \frac {m}{2}\; \varepsilon^{\mu\eta\sigma\nu}\; 
B_{\eta\sigma} \cdot D_\nu C 
+ B \cdot D^\mu C \nonumber\\
&-& (H^{\mu\nu\eta} \times K_\eta ) \cdot D_\nu C - \frac {1}{2}\; (H^{\mu\nu\eta} \times B_{\nu\eta}) \cdot C \nonumber\\
&+& \frac {i}{2}\; \partial^\mu \bar C\cdot (C \times C).  
\end{eqnarray} 
The conservation law ($\partial_\mu J^\mu_{b} = 0$) can be proven by using the following 
\begin{eqnarray}
&&D_\mu F^{\mu\nu} - \frac {m}{2}\; \varepsilon^{\mu\nu\eta\sigma} D_\mu B_{\eta\sigma} 
+ D_\mu (H^{\mu\nu\eta} \times K_\eta) 
+ \frac {1}{2} \;(H^{\nu\eta\sigma} \times B_{\eta\sigma})\nonumber\\
&& - \;\partial^\nu B - \; i \; (\partial^\nu \bar C \times C) = 0, \quad 
\partial_\mu (D^\mu C) = 0, \quad  D_\mu (\partial^\mu \bar C) = 0,\nonumber\\
&&D_\mu H^{\mu\nu\eta} - \frac {m}{2} \; \varepsilon^{\nu\eta\rho\sigma} \; F_{\rho\sigma} = 0, 
\qquad (H^{\mu\nu\eta} \times F_{\nu\eta}) = 0,
\end{eqnarray}
that emerge from the Lagrangian density (9) due to the Euler-Lagrange equations of motion. The
above conserved current leads to the derivation of the conserved charge $Q_b = \int d^3 x J^0_b$.
The latter can be succinctly expressed as
\begin{eqnarray}
Q_b &=& \int d^3 x \;\Big[ - F^{0i} \cdot D_i C + \frac {m}{2}\; 
\varepsilon^{0ijk}\; B_{jk} \cdot D_i C + B \cdot D^0 C  \nonumber\\
&-& (H^{0ij} \times K_j) \cdot D_i C - \frac {1}{2} \;(H^{0ij} \times B_{ij}) \cdot C  
+ \frac {i}{2}\; \dot{\bar C} \cdot (C \times C) \Big].
\end{eqnarray}
Using the partial integration as well as the equation of motion corresponding to the 1-form
gauge field from (12), it can be checked that the above charge can be re-expressed,
in a more compact form, as
\begin{eqnarray}
Q_b  = \int d^3 x \;\Bigl [ B \cdot D^0 C - C \cdot \partial^0 B 
- \frac{i}{2}\; \dot {\bar C} \cdot(C \times C) \Bigr ],
\end{eqnarray} 
which is exactly same in appearance as is the form of  the  BRST charge 
in the context of non-Abelian 1-form
gauge theory (see, e.g. [12]). There is a key difference, however, at the deeper
level because in (14), we have
\begin{eqnarray}
\partial^0 B =  D_i \Bigl [ F^{i0} + \frac{m}{2} \varepsilon^{0ijk} B_{jk} - (H^{0ij} \times K_j)
\Bigr ] + \frac{1}{2}(H^{0ij} \times B_{ij}) - i (\dot {\bar C} \times C),
\end{eqnarray}
which reduces\footnote{To be precise, one knows that $\partial^0 B = D_i F^{i0} -i (\dot {\bar C} \times C)$ 
in the case of the self-interacting non-Abelian 1-form gauge theory
where there is no interaction with matter fields [12].} to the case of non-Abelian 1-form gauge 
theory in the limit $B_{\mu\nu} \to 0$. This is but natural as is evident from (1). The generator (14) 
is more general than the generator (7) for the gauge transformation because it can be checked that, 
for a generic field $\Phi$, we obtain
\begin{eqnarray}
s_b \;\Phi = - \;i \;[\; \Phi,\; Q_b \;]_{\pm}, \qquad \Phi = A_0, A_i, C, \bar C, B_{ij},
\end{eqnarray}
where $(+)-$ signs, on the square bracket, stand for the bracket to be
(anti-) commutator for the generic field $\Phi$ being (fermionic)bosonic in nature.

We close this section with the remarks that the generator $Q_b$ does not generate the BRST symmetry
transformations $s_b K_\mu = - (K_\mu \times C)$ and $ s_b B_{0i} = - (B_{0i} \times C)$ because
we have not taken into account the primary constraints\footnote{The primary constraints of the 
theory are nothing but the vanishing of the canonical momenta corresponding to the 
compensating auxiliary field $K_\mu$ and the component $B_{0i}$ of the 2-form gauge field $B_{\mu\nu}$. 
On the other hand, the momenta for the field $B_{ij}$ {\it do} exist.} in the 
generalization of the Lagrangian density (1) to the BRST level in (9). Such problems do
not arise for the 1-form gauge field $A_\mu$ because we have exploited fully the usual
gauge transformations corresponding to the 1-form gauge field that are generated by the first
class constraints (that also include the primary constraint) associated with the 1-form gauge potential.

\section{Off-shell nilpotent anti-BRST symmetry \\transformations and their generator}

Corresponding to the BRST invariant Lagrangian density (9), there exists an equivalent (but coupled) 
anti-BRST invariant\footnote{The Lagrangian densities in (9) and (17) are the most general forms 
that can be obtained by exploiting the basic tenets of BRST formalism [11]. It can be 
seen that [11]
${\cal L}_B = {\cal L}_0 + s_bs_{ab} \;( \frac {1}{4}\; B^{\mu\nu} \cdot B_{\mu\nu} + \frac {i}{2} \;
A^\mu \cdot A_\mu + C \cdot  \bar C)$ and ${\cal L}_{\bar B} = {\cal L}_0 
- s_{ab}s_b \;( \frac {1}{4}\; B^{\mu\nu} \cdot B_{\mu\nu} 
+ \frac {i}{2} \; A^\mu \cdot A_\mu + C \cdot  \bar C).$ These Lagrangian densities are unique in 
the sense that the ghost number consideration and mass-dimensions (in 4D) have been taken into account [11].} 
Lagrangian density (${\cal L}_{\bar b}$)
\begin{eqnarray}
{\cal L}_{\bar b} &=& - \; \frac {1}{4} \;F^{\mu\nu} \cdot F_{\mu\nu} 
+ \; \frac {1}{12}\; H^{\mu\nu\eta} \cdot H_{\mu\nu\eta} 
+ \frac {m}{4}\;\varepsilon^{\mu\nu\eta\kappa}\; B_{\mu\nu} \cdot F_{\eta\kappa} \nonumber\\
&-& \bar B \cdot (\partial_\mu A^\mu) + \frac {1}{2}\; (B \cdot B + \bar B \cdot \bar B ) 
- i \;  D_\mu \bar C \cdot \partial^\mu C,
\end{eqnarray}
that respects the following off-shell nilpotent ($s_{ab}^2 = 0$) and anticommuting
($s_b s_{ab} + s_{ab} s_b = 0$) anti-BRST symmetry transformations\footnote{It will
be noted that the (anti-)BRST symmetry transformations ($s_b B = 0, s_b \bar B 
= - (\bar B \times C), s_{ab} \bar B = 0, s_{ab} B = - (B \times \bar C)$)
for the {\it auxiliary} fields $B$ 
and $\bar B$ in (10) and (18) have been derived by requiring the nilpotency and
anticommutativity properties of $s_{(a)b}$.} $s_{ab}$ [11]
\begin{eqnarray}
&&s_{ab} A_\mu = D_\mu \bar C, \quad s_{ab} \bar C = \frac{1}{2}\; (\bar C \times \bar C), 
\quad s_{ab} C = i\; \bar B, \quad s_{ab}  \bar B = 0, \nonumber\\
&& s_{ab} B = - (B \times \bar C), \; s_{ab} F_{\mu\nu} = - (F_{\mu\nu} \times \bar C), 
\;s_{ab} K_\mu = - (K_\mu \times \bar C), \nonumber\\
&& s_{ab} B_{\mu\nu} = - \; (B_{\mu\nu} \times \bar C), \quad s_{ab} H_{\mu\nu\eta} = - \; (H_{\mu\nu\eta} \times \bar C)
\end{eqnarray}
because $s_{ab} {\cal L}_{\bar b} = - \partial_\mu [\bar B \cdot D^\mu \bar C]$. As a consequence,
the action corresponding to the Lagrangian density ${\cal L}_{\bar b}$ remains invariant.

According to the Noether's theorem, the above continuous symmetry  transformations lead
to the following expression for conserved current
\begin{eqnarray}
J^\mu_{ab} &=& - F^{\mu\nu} \cdot D_\nu \bar C 
+ \frac {m}{2}\; \varepsilon^{\mu\nu\eta\sigma}\; 
B_{\nu\eta} \cdot D_\sigma \bar C - \bar B \cdot D^\mu \bar C \nonumber\\
&+& (H^{\mu\nu\eta} \times K_\eta) \cdot D_\nu \bar C 
- \frac {1}{2}\; (H^{\mu\nu\eta} \times B_{\nu\eta})\cdot \bar C \nonumber\\
&-& \frac {i}{2}\; \partial^\mu C \cdot (\bar C \times \bar C). 
\end{eqnarray}
The conservation law $\partial_\mu J^\mu_{ab} = 0$ can be proven by exploiting the following
Euler-Lagrange equations of motion from ${\cal L}_{\bar b}$, namely;
\begin{eqnarray}
&&D_\mu F^{\mu\nu} - \frac {m}{2}\; \varepsilon^{\mu\nu\eta\sigma} D_\mu B_{\eta\sigma} + D_\mu (H^{\mu\nu\eta} \times K_\eta) 
+ \frac {1}{2} \;(H^{\nu\eta\sigma} \times B_{\eta\sigma} )\nonumber\\
&& + \;\partial^\nu \bar B + \; i \; (\partial^\nu C \times \bar C) = 0,
\quad \partial_\mu (D^\mu \bar C) = 0, \quad  D_\mu (\partial^\mu C) = 0, \nonumber\\
&&D_\mu H^{\mu\nu\eta} - \frac {m}{2} \; \varepsilon^{\nu\eta\rho\sigma} \; F_{\rho\sigma} = 0, 
\quad (H^{\mu\nu\eta} \times F_{\nu\eta}) = 0.
\end{eqnarray}
From the equations of motion for the 1-form gauge field and (anti-)ghost fields
(cf. (12) and (20)), we obtain the celebrated CF condition $B + \bar B = - i (C \times \bar C)$
and the Lorentz gauge-fixing condition $\partial_\mu A^\mu = 0$.

The above observations show that the Lagrangian
densities (9) and (17) (i) are coupled because the above CF condition implies that
$B \cdot (\partial_\mu A^\mu) - i \partial_\mu \bar C \cdot D^\mu C =
- \bar B \cdot (\partial_\mu A^\mu) - i D_\mu \bar C \cdot \partial^\mu C$, and (ii) are equivalent in the sense that both of them respect the (anti-)BRST symmetry transformations
{\it together} because we have (besides $s_b {\cal L}_b = \partial_\mu [B \cdot D^\mu C],
s_{ab} {\cal L}_{\bar b} = - \partial_\mu [\bar B \cdot D^\mu \bar C]$)
\begin{eqnarray}
s_{ab} {\cal L}_b &=& \partial_\mu [B \cdot \partial^\mu \bar C]
- D_\mu [B + \bar B + i\; (C \times \bar C)] \cdot \partial^\mu \bar C, \nonumber\\
s_{b} {\cal L}_{\bar b} &=& - \partial_\mu [\bar B \cdot \partial^\mu  C]
+ D_\mu [B + \bar B + i \;(C \times \bar C)] \cdot \partial^\mu  C.
\end{eqnarray}
This establishes the equivalent and coupled nature of ${\cal L}_b$ and ${\cal L}_{\bar b}$.
The conserved current in (19) leads to the following conserved charge
\begin{eqnarray}   
Q_{ab}  = - \int d^3 x \;\Bigl [ \bar B \cdot D^0 \bar C 
- \bar C \cdot \partial^0 \bar B - \frac{i}{2}\; \dot C \cdot (\bar C \times \bar C) \Bigr ].
\end{eqnarray} 
The above charge generates the anti-BRST symmetry transformations for all the
relevant fields of the theory except $K_\mu$ and $B_{0i}$. 
This is due to the fact
that primary constraints corresponding to these fields have not been taken into
account in the anti-BRST invariant Lagrangian density (17).

\section{Ghost symmetry transformations and BRST algebra from symmetry generators}

It can be checked that under the following infinitesimal transformations $s_g$
\begin{eqnarray}
s_g C = + \; \Sigma \; C, \quad s_g \bar C = - \; \Sigma \;\bar C, \quad s_g [A_\mu, \; 
B_{\mu\nu}, \; K_\mu ] = 0,
\end{eqnarray}
where $\Sigma$ is a global parameter, the Lagrangian densities (9) and (17) remain invariant.
The above infinitesimal symmetry transformations are derived from the following 
explicit scale transformations
\begin{eqnarray}
C \to \;e^{\Sigma}\; C, \qquad \bar C \to  \;
e^{- \Sigma}\; \bar C, \qquad
(A_\mu, B_{\mu\nu}, K_\mu) \to  (A_\mu, B_{\mu\nu}, K_\mu),
\end{eqnarray}
where $(+)-$ signs in the above 
exponential correspond to the ghost number of a given field of the 
theory. The conserved current ($J^\mu_g$)  
and charge ($Q_g$) corresponding to the above infinitesimal transformations (23) are
\begin{eqnarray}
J^\mu_g = i \Big [ \bar C \cdot D^\mu C
- \partial^\mu \bar C \cdot C \Big], \quad
Q_g = i \int d^3 x \;\Big[ \bar C \cdot D^0 C - \dot {\bar C} \cdot C  \Big].
\end{eqnarray}
It is elementary to check that the above charge is the generator of (23).

One of the simplest ways to derive the BRST algebra is to exploit the idea of symmetry
generators amongst all the conserved charges of the theory. This can be elucidated 
in the following fashion
\begin{eqnarray}
&& s_b Q_g = - \; i\; [ Q_g, Q_b ] = - \;Q_b, \quad s_{ab} Q_g = - i \;[ Q_g, Q_{ab} ] = + \;Q_{ab}, \nonumber\\
&& s_b Q_b = - \;i \;\{ Q_b, Q_b \} = 0 \quad \Leftrightarrow \quad Q_b^2 = 0, \nonumber\\
&& s_{ab} Q_{ab} = -\;i\; \{ Q_{ab}, Q_{ab} \} = 0 \quad \Leftrightarrow \quad Q_{ab}^2 = 0, \nonumber\\
&& s_b Q_{ab} = - \;i \;\{Q_b, Q_{ab} \} = 0 \quad \Leftrightarrow \quad Q_b Q_{ab} + Q_{ab} Q_b = 0, \nonumber\\
&& s_{ab} Q_{b} = - \;i\; \{Q_{ab}, Q_{b} \} = 0 \quad \Leftrightarrow \quad
Q_b Q_{ab} + Q_{ab} Q_b = 0, etc.,
\end{eqnarray} 
which finally leads to the derivation of the well-known BRST algebra. In the above,
the anticommutativity of $Q_{(a)b}$ is proven by invoking the CF condition. 


\section{ Conclusions}

We have exploited one of the key gauge symmetries of the dynamical 4D non-Abelian 2-form
gauge theory to perform the BRST analysis. To be precise, it is the ``scalar''
gauge symmetry that has been the central symmetry of our present discussion 
and we have {\it not} even touched upon the
``vector'' gauge symmetry that is also present in the theory (cf. footnote 2).
One of the novel features of our present investigation is the fact that, even though
there exist off-shell nilpotent and anticommuting (anti-)BRST symmetry transformations
for $K_\mu$ and $B_{0i}$ fields (cf. (10), (18)), the conserved (anti-)BRST charges
(corresponding to these symmetries) are not capable of generating them. The (anti-)BRST
transformations for the former are {\it not} obtained {\it even} by the requirement of
nilpotency and anticommutativity properties of the rest of the 
(anti-)BRST symmetry transformations of our present theory.

We have provided
the possible reasons behind the existence of the above ambiguity which is, in some sense,
unique to our present gauge theory because it does not appear
in the BRST analysis of (non-)Abelian 1-form (see, e.g. [12,13]) and Abelian
2-form as well as 3-form gauge theories [8,9,14]. It should be noted that our Lagrangian densities
(9) and (17) have been obtained in their full generality by exploiting the basic tenets
of BRST formalism (see, e.g. [11] for details). However, in the BRST analysis, the
canonical momenta corresponding to $K_\mu$ and $B_{0i}$ do {\it not} appear {\it at all}
in the Lagrangian densities of the theory [11]. This is why, we have the 
presence of the above ambiguity.

It is worthwhile to mention that if we include the above constraints in the theory, then,
the  ``scalar" and ``vector" gauge symmetries mix-up together. The corresponding ``merged" (anti-)BRST symmetries turn out to be off-shell nilpotent but they do not respect the absolute anticommutativity property. Hence, these nilpotent symmetries are not proper (see [10] for details).

It would be very nice endeavor to generalize our present idea to the case of 
discussion of the ``vector'' gauge symmetries (within the framework of the
BRST formalism) that also exist in the theory. The Hamiltonian analysis
of the 4D dynamical non-Abelian 2-form gauge theory is another direction
for further investigation. These are the issues that are being pursued
at the moment and our results would be reported in our future publications.

\end{document}